# JASA ARTICLE

# A microscopic investigation of the effect of random envelope fluctuations on phoneme-in-noise perception

Alejandro Osses and Léo Varnet[a]

*Laboratoire des Systèmes Perceptifs, Département d'Études Cognitives, École Normale Supérieure, PSL University, Centre National de la Recherche Scientifique, 75005 Paris, France*

**ABSTRACT:**
In this study, we investigated the effect of specific noise realizations on the discrimination of two consonants, /b/ and /d/. For this purpose, we collected data from twelve participants, who listened to /aba/ or /ada/ embedded in one of three background noises. All noises had the same long-term spectrum but differed in the amount of random envelope fluctuations. The data were analyzed on a trial-by-trial basis using the reverse-correlation method. The results revealed that it is possible to predict the categorical responses with better-than-chance accuracy purely based on the spectro-temporal distribution of the random envelope fluctuations of the corresponding noises, without taking into account the actual targets or the signal-to-noise ratios used in the trials. The effect of the noise fluctuations explained on average 8.1% of the participants' responses in white noise, a proportion that increased up to 13.3% for noises with a larger amount of fluctuations. The estimated time-frequency weights revealed that the measured effect originated from confusions between noise fluctuations and relevant acoustic cues from the target sounds. Similar conclusions were obtained from simulations using an artificial listener. © *2024 Acoustical Society of America*.
https://doi.org/10.1121/10.0024469

## I. INTRODUCTION

Studies of speech-in-noise perception often rely on speech reception thresholds (SRTs) as a measure of speech intelligibility. An SRT is defined as the signal-to-noise ratio (SNR) at which a specific phoneme, word, or sentence score is achieved for a specific set of speech sounds [e.g., Plomp and Mimpen (1979)]. SRTs are typically obtained from many trials, obtaining an efficient overall estimate of speech intelligibility in different background noise conditions [e.g., Francart *et al.* (2011) and Stone *et al.* (2011)]. Following the rationale from some existing studies (Jürgens and Brand, 2009; Zaar and Dau, 2015), we will refer to this classical approach as providing a macroscopic view on speech perception performance. The term macroscopic refers to the fact that speech intelligibility can be efficiently characterized based on long-term characteristics of the masking sounds. Complementary to these overall insights, a microscopic view on speech perception is obtained from approaches where intelligibility is assessed on a trial-by-trial basis. A microscopic approach investigates the response variability that must be, to some extent, related to the "external variability" due to the randomness in the noise stimuli (Green, 1964). For the purposes of this study, specific noise realizations having the same long-term characteristics will be referred to as noise tokens. In this sense, a macroscopic estimate of speech intelligibility will be influenced by an "overall effect" of noise, whereas a microscopic estimate will be related to a "token-specific effect" of noise.

Macroscopic estimates of intelligibility can be interpreted in terms of energetic masking and modulation masking. The concept of energetic masking refers to a masking effect that occurs when the target sound and the (undesired) masker overlap in a set of frequency bands (French and Steinberg, 1947). In such cases, the weakest elements in the speech sounds become less audible when the SNR is decreased, until the sounds are no longer detected [see, e.g., Li *et al.* (2010), their Fig. 1]. The concept of modulation masking is similar to energetic masking but operates in the temporal modulation domain. Here, the long-term property that determines the amount of masking is the amplitude of noise envelope fluctuations within each of the analyzed frequency bands. Random envelope fluctuations have been shown to be detrimental to listening performance. As stated by Drullman (1995), they induce a "sorting problem" because the weak elements in the signal are likely to be confused with irrelevant fluctuations from the masker. Importantly, this phenomenon is present even for steady-state masking noise, such as white noise or speech-shaped noise, due to the intrinsic random envelope fluctuations present in the masker signal (Dau *et al.*, 1999). Modulation masking appears to be directly related to the amount of envelope fluctuations in the signal relative to the fluctuations in the noise (Dubbelboer and Houtgast, 2008; Jørgensen and Dau, 2011), although studies investigating modulation masking in speech-in-noise perception do not agree on the strength of this effect (Drullman, 1995; Dubbelboer and Houtgast, 2007; Noordhoek and

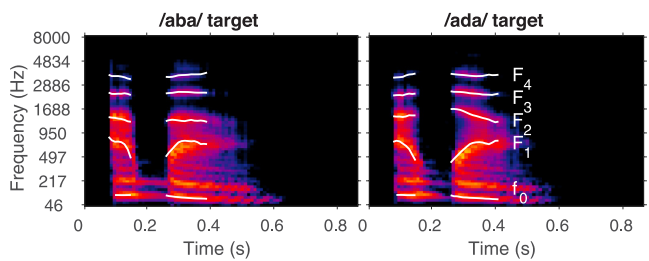

FIG. 1. (Color online) T-F representations of the two targets used in the experiment, /aba/ and /ada/. The time and frequency resolutions are the same as those used for the analysis (see Sec. II E 3 a). Lighter regions indicate higher amplitudes in a logarithmic scale. The white traces indicate the fundamental frequency ($f_0$) and formant trajectories ($F_1$–$F_4$).

Drullman, 1997; Stone et al., 2011; Stone et al., 2012). In summary, the overall impact of noise on intelligibility using the concepts of energetic and modulation masking is determined by long-term statistics of the noise masker, based on either band-level energy or the amplitude of random fluctuations, respectively. A macroscopic approach provides an efficient way to measure the impact of energetic and/or modulation masking on speech intelligibility, for example by comparing the SRT or the mean intelligibility scores between noise maskers with different long-term characteristics, measured across a large number of noise tokens [e.g., Francart et al. (2011) and Stone et al. (2011)].

However, speech intelligibility performance estimated from multiple trials may obscure information about how individual tokens are perceived [see, e.g., Singh and Allen (2012) and Zaar and Dau (2015)]. In a more general context, psychoacousticians have long been aware of the existence of a token-specific effect of noise, even if the noise tokens have the same long-term statistical properties. The studies by Pfafflin and Mathews (1966) and Pfafflin (1968) using frozen (reproducible) noise, showed that the trial-by-trial performance in a tone-in-noise detection task varies significantly over masker waveforms. In a series of two seminal studies on tone-in-noise detection, Ahumada and colleagues (1971, 1975) related the acoustic characteristics of the individual noise tokens to the corresponding responses of the participants. They used a multiple regression analysis to predict the trial-by-trial decision based on spectro-temporal representations of the noise tokens alone. The observed patterns of weights suggested that their subjects' decisions were partly related to the exact configuration of the noise token. In particular, the presence of noise energy in the spectro-temporal region of the signal led to an increase in signal-present responses. Conversely, noise tokens with more energy in the regions preceding the signal and surrounding it in frequency yielded more signal-absent responses. Following a similar approach, Varnet and Lorenzi (2022) showed that the exact temporal distribution of random envelope fluctuations in a trial has a systematic influence on the detection of an amplitude-modulated target. More specifically, these fluctuations can bias the participant's response towards perceiving a 4-Hz modulation (or not) depending on the token-specific configuration.

In speech perception, however, demonstrations of the token-specific effect of noise are more seldom. One example is provided by Zaar and Dau (2015), who performed a trial-by-trial analysis using a fixed set of frozen noise tokens. Their participants had to identify consonant-vowel words embedded in white noise at six SNRs. The authors were particularly interested in assessing the effect of various sources of variability in the task. They found that a 100-ms temporal shift of the masking noise waveform could induce a significant perceptual effect, that was well above the assessed within-listener variability. Thus, a given speech utterance was found to be either more or less robust, eliciting a different pattern of confusions, depending on whether the utterance was presented along one specific noise token or its time-shifted version [see also Cooke (2009)]. In their interpretation of this finding, Zaar and Dau (2015) noted that "[…] the common assumption in various previous studies of an invariance of consonant perception across steady-state noise realizations cannot be supported by the present study. In fact, the results obtained here suggest that the interaction between a given speech token and the spectro-temporal details of the "steady-state" masking noise waveform matter in the context of microscopic consonant perception. When analyzing responses obtained with individual speech tokens, averaging responses across noise realizations thus appears problematic" (p. 1263). This conclusion is supported by the results in two of our previous studies on phoneme discrimination in white noise (Varnet et al., 2013; Varnet et al., 2015). Similar to the method adopted by Ahumada et al. (1975), the microscopic approach used in these studies was based on a trial-by-trial statistical analysis that relates the random envelope fluctuations of the noise with the corresponding response of the listener. The outcome of this analysis is a time-frequency (T-F) matrix of perceptual weights, named auditory classification image (ACI), which highlights the T-F regions of the stimulus where an increase in random envelope fluctuations induces a systematic bias in the listener's phonetic decision. The measured ACIs revealed a significant pattern of weights, therefore confirming the role of the token-specific effect of noise on phoneme perception.

The above set of findings based on microscopic approaches (frozen noise or ACI) suggests that the overall effect of noise on speech perception must be further complemented by the assessment of a token-specific effect. However, the strength of this effect remains a debated question. For example, Régnier and Allen (2008) measured the variance caused by the masker in an auditory representation of the phoneme /t/ embedded in white and speech-shaped noises, concluding that—at least for this noise-robust phoneme—the SNR range in which noise and speech information interacted was in fact very limited and that different noise tokens should not significantly impact phonetic judgements. Contrary to these conclusions, Zaar and Dau (2015) estimated a significant effect of the variability in the background noises on their listeners' decisions, although this effect was found to be smaller than the effect induced by the variability in the speech sounds. Support for these observations can also be found in the preliminary study by Cooke (2009), who estimated a small effect of noise on word confusions, finding consistent confusions in 7% of the noisy speech stimuli for a majority of his participants.

The present study investigates the token-specific effect of random envelope fluctuations introduced by noise[1] on the perception of speech acoustic cues in a consonant-in-noise discrimination task using nonsense words of the structure vowel-consonant-vowel. More specifically, the task is tested using an /aba/ and an /ada/ utterance presented in three types of background noises. All noises have the same flat long-term averaged spectrum but differ in the amount of random envelope fluctuations: A Gaussian white noise, a noise with a band limited modulation spectrum, or a noise with randomly imposed bursts of energy. The overall effect of noise on the participants' performance is quantified based on the SNR required to successfully discriminate the target sounds with a 70.7% correct response rate. Additionally, the token-specific effect of noise on performance is quantified by means of an analysis of the exact trial-by-trial random envelope fluctuations that is inherent to the reverse-correlation method used to derive ACIs.

In this context, our working hypotheses (H1–H4) are as follows:

*H1*: Using the derived ACIs and the specific set of noise tokens used during the experiments, we can predict the response ("aba" or "ada") of each participant with an accuracy that is significantly above chance. The prediction performance metrics will provide us with a measure of the strength of the token-specific effect in this speech-in-noise task.
*H2*: Noise conditions differing only with respect to their modulation content will induce a different token-specific effect. More specifically, for a given overall performance level, noises with a larger amount of random envelope fluctuations will yield a higher ACI prediction performance.
*H3*: The ACIs will be globally similar for all individuals and conditions. Not only the token-specific effect should be measurable in each listener, but it should impact the same cues for every participant, as the individual listening strategies should be globally similar.
*H4*: Noises with a larger amount of random envelope fluctuations should yield better predictions when ACIs are derived using simulated responses from an artificial listener. This hypothesis can be seen as a way to jointly test H1–H3 based on a decision strategy that integrates signal-driven (bottom-up) cues as an ideal observer would do. In other words, the artificial listener is used as a baseline for performance (Green and Swets, 1966).

All hypotheses (H1–H4) were preregistered before data collection (Osses and Varnet, 2022c).

## II. MATERIALS AND METHODS

All stimuli and procedures were preregistered (Osses and Varnet, 2024) and can be reproduced with the fastACI toolbox (Osses and Varnet, 2023), which in the following we refer to as "the toolbox."

### A. Stimuli

#### 1. Target sounds

We used two male speech utterances from speaker S43M taken from the OLLO speech corpus (Meyer *et al.*, 2010) (/aba/: S43M_L007_V6_M1_N2.wav; /ada/: S43M_L001_V6_M1_N1.wav). We preprocessed these speech samples to align the time position of the vowel-consonant transitions, to equalize their energy per syllable, and to have the same waveform duration (Osses *et al.*, 2022a). The stored sounds have a duration of 0.86 s, a sampling frequency $f_s$ of 16 kHz, and an overall level of 65 dB sound pressure level (SPL). The time-frequency (T-F) representation of the stored sounds is presented in Fig. 1, together with their fundamental-frequency ($f_0$) and formant ($F_1$–$F_4$) trajectories.

#### 2. Background noises

Three types of background noise conditions were tested. These conditions were chosen to include a stationary noise condition using white noises and two additional noises with stronger envelope fluctuations below 35 Hz, that correspond to non-stationary noise conditions. To increase the low-frequency fluctuations in the envelope domain, we designed an algorithm to generate a white noise with superimposed Gaussian bumps and an algorithm to generate noises with limited modulation power spectrum (MPS). We abbreviate these two types of noises as bump and MPS noises, respectively. The noises were generated at an $f_s$ of 16 kHz with a duration of 0.86 s and an overall level of 65 dB SPL. The noises were subsequently gated on and off with 75-ms raised-cosine ramps before being stored on disk.

The acoustic characteristics of the noises are shown in Fig. 2 and the algorithm details are given in the next paragraphs. For each noise we show the T-F representation of one arbitrarily chosen noise realization [Fig. 2(a)], followed by an acoustic analysis of the noises derived from 1000 noise realizations [Figs. 2(b) and 2(c)] using (1) critical-band levels within 1 equivalent rectangular bandwidth (ERB) for bands centered between 87 Hz (or 3 $ERB_N$) and 7819 Hz (or 33 $ERB_N$) in Fig. 2(b) and (2) assessing the broadband envelope spectrum obtained from the absolute value of the Hilbert envelope in Fig. 2(c), referenced to their mean (DC) value, equal to 66.2 dB for all noises [0 dB re. max in Fig. 2(c)].

The generated white noises had a spectrum level of 26 dB/Hz with an effective bandwidth between 0 and 8000 Hz, resulting in critical-band levels between 40.7 dB and 56.2 dB [Fig. 2(b)]. The envelope spectrum [Fig. 2(c)] was approximately constant with a median amplitude of –44 dB re. max, although a theoretical monotonic decrease up to $f_s/2$ is expected (Dau *et al.*, 1999). This decrease is not visible due to the 0–60 Hz limit of the abscissa.

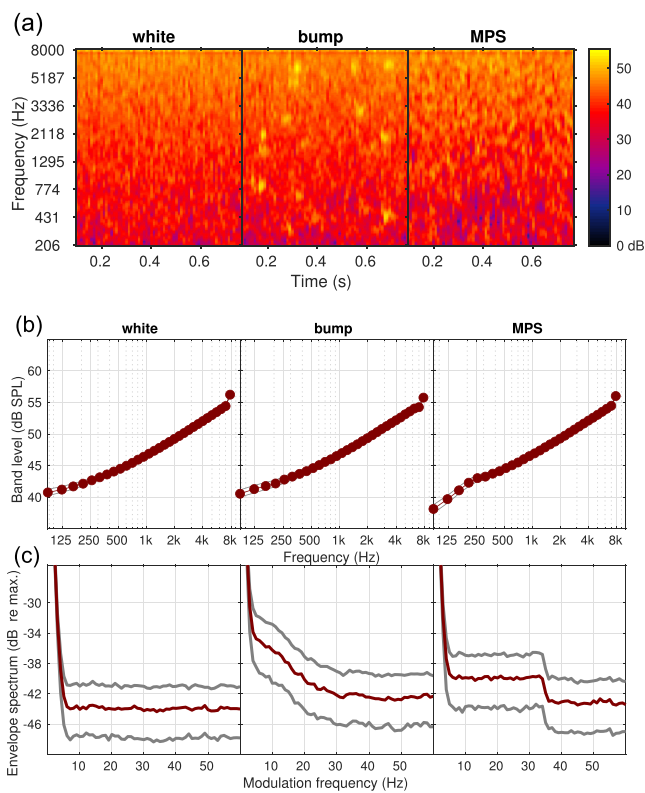

FIG. 2. (Color online) Summary of the acoustic characteristics of white (left), bump (middle), and MPS noises (right). (a) T-F representation of one arbitrarily chosen noise representation shown as magnitudes in dB. (b) Critical-band levels within 1-ERB wide filters. (c) Envelope spectrum referenced to the mean (DC) value. More details are given in the text. In panels (b) and (c), the gray curves indicate the percentiles 25 and 75 of the corresponding estimate.

The bump noises were generated using an algorithm similar to that described by Varnet et al. (2019). Each noise token was generated starting from a Gaussian noise to which bumps were superimposed. The bumps are regions of excitation that have a Gaussian shape defined by a temporal width of $\sigma_t = 0.02$ s and a spectral width of $\sigma_f = 0.5$ ERB, amplified by up to 10 dB. The time and frequency locations of the bumps were randomly spread across the entire duration of the initial Gaussian noise and through the whole T-F space, i.e., between 80 and 7158 Hz (i.e., 1 $ERB_N$ below 8000 Hz). Each waveform contained 30 newly drawn Gaussian bumps. The generated bump noises had critical-band levels between 40.5 and 55.8 dB [Fig. 2(b)]. The envelope spectrum [Fig. 2(c)] had a triangular shape going from an amplitude of −34.8 dB re. max at $f_{mod} = 3$ Hz down to −42.7 dB re. max at $f_{mod} = 31.1$ Hz with an approximately constant spectrum thereafter (median amplitude of −42.5 dB re. max).

Finally, the MPS noises were generated by limiting their spectrum in the modulation frequency domain using a set of temporal and spectral rate cut-off frequencies. We chose a temporal cut-off of 35 Hz and a spectral cut-off of 10 cycles/Hz, based on the study by Elliott and Theunissen (2009) and some pilot tests. The MPS bandwidth was limited using the phase reconstruction approach from the PhaseRet toolbox (Průša, 2017). Inspired by Venezia et al. (2016), we first generated a white noise that is multiplied in the MPS domain by a low-pass envelope with the desired characteristics defined by the temporal and spectral cut-off frequencies. The MPS-limited representation was then converted back to a time-domain waveform and stored on disk. The generated MPS noises had band levels between 38.1 and 56.0 dB [Fig. 2(b)]. The envelope spectrum [Fig. 2(c)] had a constant value of −39.9 dB re. max starting after the mean (DC) component at 0 Hz and up to about 35 Hz, an amplitude that decreases to a constant value of −43.2 dB re. max thereafter.

In summary, all noises originate from white noises with or without emphasized envelope fluctuations, sharing nearly the same long-term spectral content [Fig. 2(b)]. However, the noises have a different amount and distribution of random envelope fluctuations in the modulation-frequency domain [Fig. 2(c)]. White noises have low envelope fluctuations, MPS noises have rectangular-shaped low-pass envelope fluctuations (cut-off $f_{mod} = 35$ Hz), and bump noises have triangular-shaped low-pass envelope fluctuations (for $f_{mod} < 31.1$ Hz).

For each participant a new set of 4000 noises with a level of 65 dB SPL was generated, resulting in 36 sets of noises (12 participants × 3 noise conditions). The simulated participant, the artificial listener (Sec. II D), was tested on the same 36 sets of experimental noises. All noise waveforms were stored and can either be retrieved from Zenodo (Osses and Varnet, 2022b) or be reconstructed using the toolbox (see supplementary material Sec. II).

### 3. Noisy trials

The target sounds were first adjusted in level, depending on the trial SNR, and were then arithmetically added to the corresponding noise, following the staircase rule detailed later in this section. Before the trial was administered to the listeners, an additional but small variation in the total presentation level (level roving) between −2.5 and +2.5 dB (continuous range, uniform distribution) was applied to partly discourage the use of loudness cues during the experiment. For the simulations using the artificial listener that were run with all 36 noise data sets, the same order of trials and level roving were applied as used for each participant.

### B. Apparatus and procedure

The experiments were conducted in one of the two doubled-walled soundproof booths located at our group facilities (LSP, ENS Paris). The experiment utilized a within-subject design. In each trial, the nonsense words /aba/ or /ada/ were presented diotically via Sennheiser HD 650 circumaural headphones (Sennheiser, Wedemark, Germany) in one of three background noises. The task of the participant was to indicate whether they heard "aba" or "ada" by pressing "1" or "2" on the computer keyboard, respectively.

#### 1. Experimental protocol

The participants completed 4000 trials using each noise type (total of 12 000 trials), organized in thirty test blocks of 400 trials. A block contained approximately 200 trials of each target sound and its evaluation took between 12 and 15 min. The order of the test blocks was pseudo-randomized with the only constraint that the three noise conditions were presented in permuted order in the first three blocks. Subsequently, all blocks were randomly assigned. Each participant required five or six two-hours sessions to complete the experiment.

For each test block only one type of noise was evaluated and one independent adaptive track was measured: After a correct or incorrect response, the level of the target word in the subsequent trial was decreased or increased, respectively, following a one-up one-down weighted adaptation rule (Kaernbach, 1991). We used up- and down-steps in a ratio of 2.41 to 1 that lead to a target a score 70.7% according to Kaernbach (1991), his Eq. (1). Participants received feedback on the correctness of the trial. Furthermore, they were explicitly instructed to minimize their response bias as much as possible with a warning message displayed on screen when the response ratio was higher than 60% or lower than 40%.

For each trial, we stored the participant's response, the corresponding SNR, the target actually presented, and the exact waveform of the noise. After completing each block, participants were encouraged to take a short break.

#### 2. Training session

Before the first test block, the participants completed a short training to make sure that they correctly understood the task. This training session was similar to the main experiment except that participants were able to repeat the noisy speech stimuli or to listen to /aba/ or /ada/ samples in silence. The training results were excluded from any further analysis.

### C. Participants

Twelve participants (S01–S12) aged between 22 and 43 years old (4 females, 8 males) took part in our study, with eight of them being native French speakers. The participants were volunteers with self-reported normal hearing. Average audiometric thresholds and other details about the participants are given in supplementary material Sec. I. The participants provided their written informed consent prior to the data collection and were paid for their contribution.

The total number of $N = 12$ participants exceeds the preregistered number of participants ($N = 10$). By the time we completed the data collection for $N = 10$, twelve participants had been recruited, and we therefore decided not to interrupt the data collection for the last two subjects (S06, S12). The results reported in this study are based on all twelve participants. The same analysis using the preregistered sample size can be found in supplementary material Sec. III B and yielded very similar results.

### D. Simulated participant: The artificial listener

In addition to the experimental data collection, we used an auditory model to simulate the performance of an average normal-hearing listener who uses a fixed decision criterion to compare sounds. This "artificial listener" assesses the internal representations of each sound using signal-driven (bottom-up) information based on a modulation-filter-bank approach (Dau et al., 1997). The internal representations were subsequently compared using a (top-down) decision back-end based on template matching, with two stored templates, one for each target sound (Osses and Kohlrausch, 2021). The artificial listener was treated as an additional participant, meaning that its results were subjected to the same data analysis as applied to the experimental data.

The auditory model is composed of a front-end and a back-end processing using default parameters in our toolbox. In short, the front-end processing is very similar to the model described by Osses and Kohlrausch (2021), except that the middle-ear module is implemented as a linear phase filter and that the modulation filter bank uses a Q factor of 1. The two templates were derived at a supra-threshold SNR of −6 dB where each target was embedded and subsequently averaged across 100 newly generated white noise realizations. The exact supra-threshold SNR and the number of averaged realizations were arbitrary choices.[2] This fixed white-noise template was used for the simulations in all three noise conditions. The trial-by-trial decision was based on a template-matching where a decision bias was introduced to allow the model to balance the number of "aba" and "ada" choices (Osses and Varnet, 2021). All details about the model configuration and the decision scheme can be found in the supplementary material Sec. IV.

### E. Data analysis

The experimental data collection resulted in 36 staircases for twelve participants in the three noise conditions and 36 staircases for the simulations with the artificial listener. In this section we describe the analyses that were applied to the experimental trials to obtain the direct behavioral results and to derive the individual ACIs.

#### 1. Preregistered data exclusion criteria

For a more efficient data processing, the ACI method requires a minimization of response biases. Next to the explicit instruction to balance "aba" and "ada" choices (Sec. II B 1), we preregistered two criteria for trial exclusion.

The first criterion is related to the exclusion of all starting trials of each test block before reaching the fourth turning point or reversal. Those trials correspond to the so-called approaching phase of the staircase procedure, where the adjustable parameter, the SNR, is considered to be at a supra-threshold level with a percentage correct that is well above the target 70.7%. The fourth reversal was considered to be the starting point of the measuring phase of the staircase.

The second criterion is an explicit control of the balance between "aba" and "ada" responses in our dataset. During the data processing, the responses of the target sound that obtained more preferences were sorted in increasing SNR. Subsequently the trials with most extreme values (minima or maxima) were discarded until the same number of "aba"-"ada" preferences was achieved. In other words, if a participant indicated "aba" 53% of the times and "ada" 47% of the times, the trial exclusion was applied to the "aba" trials.

### 2. Measures of behavioral performance

The listeners' performance in the different noise conditions was assessed using a number of measures derived from the trial SNRs. The percentage of correct responses and SNR thresholds were obtained for each block of 400 trials, after data exclusion (Sec. II E 1). Then, the rate of correct responses in /aba/-trials and in /ada/-trials were expressed in histograms using SNR bins of 1 dB. Note that these values correspond to the rate of hit and correct rejection if we arbitrarily identify /ada/- and /aba/-trials as target-present and target-absent trials, respectively. Finally, using the same 1-dB wide bins, the classical discriminability index ($d'$) and criterion ($c$) metrics were obtained from the hit, false alarm, correct rejection, and miss rates (Harvey, 2004) as a function of SNR.

The behavioral measures $d'$, $c$, and the block-by-block SNR threshold were tested for a group-level effect using a mixed analysis of variance (ANOVA) with two fixed factors, block number and noise condition. Participants were treated as a random effect, meaning that differences in baseline performance for individual listeners were taken into account. This analysis was run to test for learning effects. A second mixed ANOVA with two fixed factors, SNR and noise condition, was run to confirm the effect of SNR on $d'$. Similarly, a mixed ANOVA was also run on the criterion $c$.

### 3. Auditory classification images (ACIs)

*a. Time-frequency (T-F) representations.* Following the same rationale as in previous studies, the ACIs were derived and interpreted in a T-F space (Osses and Varnet, 2021; Varnet et al., 2013; Varnet et al., 2015). Here, we chose to use a Gammatone-based representation rather than a spectrogram. The 0.86-s long monaural noises were decomposed into 64 bands equally spaced in the ERB-number ($ERB_N$) scale (Glasberg and Moore, 1990) between 45.8 Hz (1.69 $ERB_N$) and 8000 Hz (33.19 $ERB_N$), spaced at 0.5 ERB. The filters had a width of 1 ERB, resulting in a 50% overlap. The 64 band-passed signals were then low-pass filtered using a Butterworth filter ($f_{cut-off} = 770$ Hz, fifth order), which roughly simulates the inner-hair-cell envelope extraction processing [see, e.g., Osses et al. (2022b), their Sec. 2.4]. Finally, one estimate every 0.01 s (amplitude mean) was obtained for each of the frequency bands along the time dimension resulting in a final T-F noise representation stored in a 86-by-64 matrix. We denote the T-F representation of the noise presented to participant $k$ in trial $i$ as $\underline{\underline{N}}_{k,i}$, while $\underline{N}_{k,i}$ refers to the vectorization of this matrix. In the following, we use the same formalism to refer to the ACI in its matrix form ($\underline{\underline{ACI}}$, 86-by-64) or vector form ($\underline{ACI}$, 5504-by-1).

*b. Generalized linear model.* The core principle of the ACI approach is to assess how the random envelope fluctuations in the stimulus ($\underline{\underline{N}}_{k,i}$) affect the behavioral response of the participant (denoted $r_{k,i}$) on a trial-by-trial basis. For this purpose, we relied on a stimulus-response transformation based on a generalized linear model (GLM) to produce a T-F matrix of decision weights (Varnet et al., 2013; Varnet et al., 2015). As the objective of our study was to isolate the effect of random envelope fluctuations on phoneme perception, the GLM did not include any complementary predictor like the target actually presented or the SNR. We define the vectorized ACI for participant $k$, $\underline{ACI}_k$, such that

$$P(r_{k,i} = \text{"aba"}) = \Phi(\underline{N}_{k,i}^T \cdot \underline{ACI}_k + c_k), \quad (1)$$

where $P(r_{k,i} = \text{"aba"}) = 1 - P(r_{k,i} = \text{"aba"})$ is the predicted probability of choosing "aba" and $\Phi(x)$ stands for the sigmoid function $\Phi(x) = 1/(1 + e^{-x})$. Equation (1) relates the specific content of the noise in trial $i$ to the response given by the participant, with $\underline{ACI}_k$ and $c_k$ being the GLM parameters that need to be fitted to each participant's data. The $\underline{ACI}_k$ in Eq. (1) is expressed as a vector of perceptual weights, with each element corresponding to one T-F point of the noise representation $\underline{N}_{k,i}$. Therefore, they are more easily interpreted as a matrix $\underline{\underline{ACI}}_k$ that has the same size as the $\underline{\underline{N}}_{k,i}$ matrix. The parameter $c_k$ corresponds to the fitted intercept value that indicates the overall bias of the participant towards one response or the other.

*c. Sparseness prior in a Gaussian pyramid basis.* An ACI is typically composed of many non-zero weights. However, these weights are often grouped into positive or negative clusters matching the location of acoustic cues in the targets, while the rest of the T-F space is close to zero. Therefore, a more compact way of describing an ACI would be as a linear combination of Gaussian-shaped elements centered at different T-F locations, such as the ones shown in Fig. 3. Here, formulating the problem in a space where ACIs can be expressed with a limited number of coefficients allows us to enforce "sparse" solutions, that is, ACIs that are non-zero only in a few localized T-F regions.

The Gaussian pyramid consists of four successive levels (1 to 4) corresponding to decreasing T-F resolutions. Each level is composed of Gaussian elements of the same width (standard deviation = 1, 2, 3, or 4 bins, respectively) and spaced every 1, 2, 3, or 4 bins. This means that the first level is not subsampled in contrast to the gradually more subsampled levels 2 to 4.[3] The coefficients of the Gaussian elements from all four levels are normalized to have a norm equal to 1, vectorized, and stored into a single matrix $\underline{\underline{B}}$. An ACI is then expressed as a linear combination of Gaussian elements,

$$\underline{ACI}_k = \underline{\underline{B}} \cdot \underline{\beta}_k, \quad (2)$$

which represents a change of basis that relates the coordinates of the ACI in the new multi-resolution Gaussian-pyramid space ($\underline{\beta}_k$), to its coordinates in the T-F space. By replacing $\underline{\underline{ACI}}_k$ in Eq. (1) we obtain

$$P(r_{k,i} = \text{``aba''}) = \Phi(\underline{N}_{k,i}^T \cdot \underline{\underline{B}} \cdot \underline{\beta}_k + c_k). \quad (3)$$

Instead of estimating the ACI directly in the T-F space, using Eq. (1), we actually estimated the $\underline{\beta}_k$ coefficients of Eq. (3). Although Eqs. (1) and (3) are mathematically equivalent, the latter allows the prior assumption about the simplicity of the ACI to be expressed by looking for as few non-zero $\underline{\beta}_k$ coefficients as possible (Mineault et al., 2009). In statistical terms, this is achieved by penalizing the classic maximum-likelihood estimator with a L1-regularized (lasso) regression approach, which enforces sparse solutions. The weight and bias for each participant ($\underline{\beta}_k$ and $c_k$) were fitted individually with a lasso regression, then transformed back into the T-F space using Eq. (2) to obtain the final $\underline{\underline{ACI}}_k$. Following the standard lasso procedure, the hyperparameter $\lambda$ that controls the strength of the regularization was selected to minimize the deviance through a 10-fold cross-validation approach. We tested twenty plausible $\lambda$ values logarithmically spaced between $1.1 \times 10^{-3}$ and 0.1, with larger values enforcing more sparse candidates, as shown in supplementary material Fig. 7 [see also Varnet et al. (2015), their Fig. 2]. The choice of this range of values ensured that the lowest amount of regularization was low enough to produce a very noisy ACI, while the highest amount of regularization was high enough to produce a flat ACI. A flat or "null" ACI, only contains weights that are equal to zero, meaning that an ACI prediction is only defined by the bias $c_k$. This statistical-fitting procedure is the same as we have used in our latest studies (Osses and Varnet, 2021, 2022a).

### 4. Out-of-sample prediction

*a. Performance metrics.* Following the standard hyperparameter selection procedure for GLMs [e.g., Mineault et al. (2009), Varnet et al. (2015), and Wood (2017)], the out-of-sample predictive performance of the fitted ACIs was assessed during the 10-fold cross-validation using the cross-validated deviance. To allow a direct comparison between different ACIs, that differ in the exact number of test trials due to the criteria for trial exclusion (see Sec. II E 1), we report the cross-validated deviance per trial (CVD$_t$).

We adopted a second complementary metric that we defined as prediction accuracy (PA). PA is a "noisier" but more intuitive measure of prediction performance that is assessed as the coincidence between predicted and actual responses. PA relates the predicted and actual responses, expressing "aba" (or "ada") predictions when /aba/ (or /ada/) was actually chosen by the participant. Assuming that a probability $P$ equal to or above 0.5 in Eq. (1) [or Eq. (3)] would be related to a predicted choice of "aba," the PA metric can be formalized as

$$\text{PA}_i = \begin{cases} 1 & \text{if } P(r_{k,i} = \text{``aba''}|\text{/aba/presented}) \geq 0.5, \\ 1 & \text{if } P(r_{k,i} = \text{``aba''}|\text{/ada/presented}) < 0.5, \\ 0 & \text{otherwise.} \end{cases} \quad (4)$$

This metric was averaged across trials and expressed as a percentage. The metric can adopt values between chance ($\sim$50%) and 100%.

To facilitate the interpretation of the previous performance metrics, we define the deviance-per-trial benefit $\Delta$CVD$_t$ and the percent accuracy benefit $\Delta$PA as the difference between prediction performance using the optimal ACI [Eq. (3)] and that of the corresponding null ACI. For $\Delta$PA we further scaled the metric by $1/(1 - \text{PA}_{null})$. This way, PA values that can range between $\sim$50% and 100% are mapped to $\Delta$PA values between 0% and 100%.

Given that lower $\Delta$CVD$_t$ values indicate better predictions, for individual evaluations we provide one-sided 95% confidence intervals, obtained as 1.64 times the standard error of the mean (SEM), reporting a significant benefit if the confidence interval is below zero. For the group evaluation, mean $\Delta$CVD$_t$ values across folds were obtained for each participant and the significance was assessed at the group level following a similar criterion.

With the $\Delta$PA metric, the benefit of using the optimal ACI with respect to the null ACI is expected to increase. As a reference, we show performance boundaries at 2.6% or 4.78% for evaluations using all experimental trials or incorrect trials only, respectively.[4]

For the purposes of this study, the metrics of prediction accuracy are not only a way to validate ACIs, but they also provide a proxy for the size of the token-specific effect on phoneme perception with better values when more of the participant confusions are due to random envelope fluctuations. Strictly speaking, the predictability gives us a lower boundary of the token-specific effect, as responses that are correctly predicted using a model that by design is only based on random envelope fluctuations—as it is the case here for the T-F representations transformed using the fitted GLMs—must be caused by those random envelope fluctuations. On the contrary, incorrect predictions could either be due to token-specific effects that are not accounted for in our GLM approach, such as interactions between separate T-F regions or effects where non-linear processes are involved, or to other causes. For these reasons, we also report the performance metrics only using data from incorrect trials. In this case, the metrics are labeled as $\Delta CVD_{t,inc}$ and $\Delta PA_{inc}$, respectively.

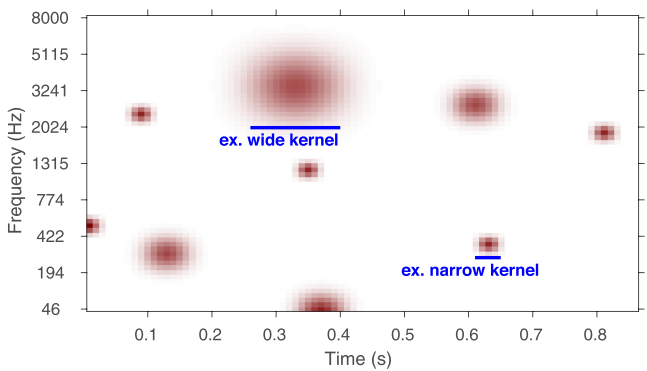

FIG. 3. (Color online) Illustrative example of a nine Gaussian basis element used in the pyramid decomposition represented in a T-F space, as used for the ACIs. Thanks to the multi-resolution matrices of the decomposition method, narrower and wider cues can be extracted from the input noise matrix $S_{k,i}$. Two stereotypical basis elements are indicated by the blue lines.

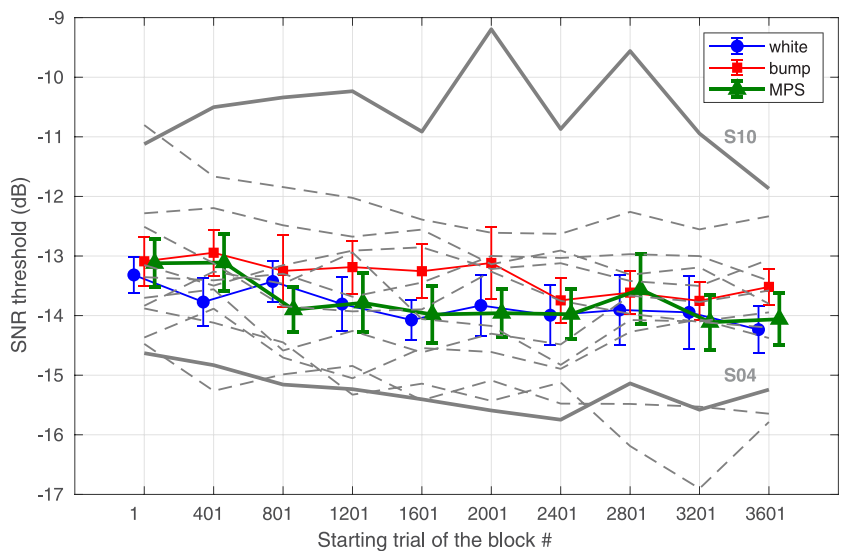

FIG. 4. (Color online) Mean SNR thresholds for the group in blocks of 400 trials for each of the three masker conditions. We show the individual thresholds for the twelve participants averaged across conditions (gray dashed lines), emphasizing the overall thresholds of the participants with lowest and highest values (thick gray continuous lines). The error bars indicate one standard error of the mean (SEM).

*b. Cross-predictions.* The selected performance metrics measure the ability of the ACI, fitted on a subset of the participant's data, to predict unseen data from the same participant in the same condition. Complementary to these "auto-predictions," we also derived "cross-predictions," where the fitted ACI was evaluated on a test set extracted from a different participant or a different condition. We assessed two types of cross-predictions: (1) within participant but between conditions and (2) between participants but within conditions. While the auto-predictions were used to assess the goodness-of-fit of the obtained ACIs, the cross-predictions were used to evaluate the similarity between listening strategies across participants or across maskers. To test the significance of the cross-predictions, two ACIs were considered as "similar" if we could exchange them and reach a significantly better-than-chance prediction accuracy, i.e., with newly obtained $\Delta CVD_t$ being significant according to the criteria defined above. We also report the correlation across ACIs, but this complementary analysis is only presented in supplementary material Sec. III.

## III. RESULTS

For each participant, the experimental data collection was completed across different days, mostly requiring five two-hour sessions. All the recruited participants ($N = 12$) were able to complete the task, although participant S10 showed highly variable results with scores that were clearly below the group average.

### A. Measures of behavioral performance

In the course of the experiment, the level of the speech target (/aba/ or /ada/) was adapted using a one-up one-down weighted adaptation rule that targeted a 70.7% of correct responses (see Sec. II B 1). In practice, after excluding the approaching phase of the staircases (see Sec. II E 1), the exact percentage of correct responses averaged across noise conditions and test blocks ranged between 71.0% (S10) and 71.6% (S03), with session-by-session scores between 69.5% and 74.2%.

To provide an overview of the participants' performance, the obtained SNR thresholds as a function of test block are shown in Fig. 4. In this figure, we show the overall performance (averaged across participants) for white- (blue), bump- (red), and MPS-noise conditions (green). Additionally, the SNRs for each participant were averaged per block for each noise condition obtaining twelve gray traces, also shown in Fig. 4. The thresholds of participants S04 and S10 are shown in solid lines and correspond to the (overall) best and worst performing participants in the task, respectively. The SNR evolution given by the gray traces suggests that there was a small learning effect during the course of the experiment, with slightly better (lower) thresholds in the last test blocks. This small learning effect was confirmed by our first ANOVA, with factors masker and test block (see Sec. II E 2). Both factors were found to have a significant effect on the obtained SNR thresholds with $F(2,345) = 15.87$, $p < 0.001$ and $F(1,345) = 36.63$, $p < 0.001$, respectively. A *post hoc* analysis revealed that the effect of masker type was in fact due to a difference in the bump-noise condition compared to the other two, while SNR thresholds for white noise and MPS noise were not significantly different.

In order to measure the effect of speech level on performance, trial-by-trial responses were converted to mean scores, $d'$, and criterion values ($c$) as a function of SNR. These metrics are shown in Fig. 5. We then ran the two other ANOVAs (Sec. II E 2). For these tests we only used the data for the SNR bins centered between –16 and –12 dB, where data for all participants in all conditions had been obtained. The ANOVAs supported a significant effect of the factors masker and SNR on $d'$ [masker: $F(2,165) = 10.39$, $p < 0.001$; SNR: $F(1,165) = 1017.65$, $p < 0.001$], while only the factor SNR had a significant effect on criterion [masker: $F(2,165) = 1.75$, $p = 0.178$; SNR: $F(1,165) = 9.77$, $p = 0.002$]. According to a *post hoc* test, the effect of masker type on $d'$ was due to a difference in the white-noise condition compared to the other two.

## B. ACIs

The ACIs that were derived from the collected data are shown in Fig. 6 using white, bump, and MPS noise maskers. Panels (A)–(F) contain the individual ACIs, while a group average is shown in the bottom-most panels [Fig. 6(g)–6(i)]. Overall, the individual ACIs bear some similarities, although for two participants (S09 and S10) the hyperparameter selection in white noise did not yield a minimum, resulting in a null ACI (dashed pink boxes in Fig. 6). Large weights were found at $t \approx 0.3$ s, the time of the onset of the second syllable, which are more clearly visible in the group ACIs. More specifically, we found a clear pattern of positive (red) and negative (blue) weights matching the location of the $F_1$ and $F_2$ onsets of the /aba/ and /ada/ sounds. Additionally, in a subset of ACIs, weak but consistent perceptual weights were also found around the time of the first-syllable $F_2$ offset (e.g., in the ACIs of S03, all conditions at $t \approx 0.25$ s), or near the release burst of the plosive consonant at around 8 kHz (e.g., ACI of S07, white-noise condition at $t \approx 0.6$ s).

The group ACIs were obtained as the arithmetic average of all non-normalized individual ACIs for white [Fig. 6(g)], bump [Fig. 6(h)], and MPS noises [Fig. 6(i)] and are only shown for visualization purposes. In these panels we superimposed the $f_0$ and formant trajectories of /aba/ and /ada/. The group ACIs were not normalized to emphasize the fact that the (blue and red) weights have different limits for the different noises.

## C. Out-of-sample prediction accuracy

**Auto-predictions at the individual level**: The out-of-sample metrics of prediction accuracy, $\Delta CVD_t$ and $\Delta PA$, at the individual and group level are shown in Fig. 7, where the metrics for the individual ACIs are shown as open or gray diamond markers.

The results based on $\Delta CVD_t$ [Fig. 7(a)] show that 27 ACIs out of 36 yielded predictions that were significantly higher than chance. Those significant estimates are marked in gray in Fig. 7(a). From the remaining nine non-significant ACIs, two corresponded to the conditions where a null ACI was obtained (dashed pink boxes in Fig. 6), which, by definition, are related to performance metrics equal to zero.

For the analysis of incorrect trials, the results are shown in Figs. 7(b) and 7(d) for $\Delta CVD_{t,inc}$ and $\Delta PA_{inc}$, respectively. Although the improvement in $\Delta CVD_{t,inc}$ was rather small [compare Fig. 7(b) with 7(a)], there was a systematic improvement of $\Delta PA_{inc}$ values [compare Fig. 7(d) with 7(a)].

**Auto-predictions at the group level**: At the group level, white noise yielded a smaller prediction performance compared to the bump and MPS noise conditions with $\Delta CVD_t$ values of $-0.657 \times 10^{-2}$, $-1.490 \times 10^{-2}$, and $-1.207 \times 10^{-2}$ [filled maskers in Fig. 7(a)] and $\Delta PA$ values of 8.1%, 13.3%, and 11.8%, respectively [filled maskers in Fig. 7(c)]. Additionally, a significant effect was found for the factor "masker" in a mixed ANOVA on $\Delta CVD_t$ [$F(2,22) = 7.73$, $p = 0.003$].

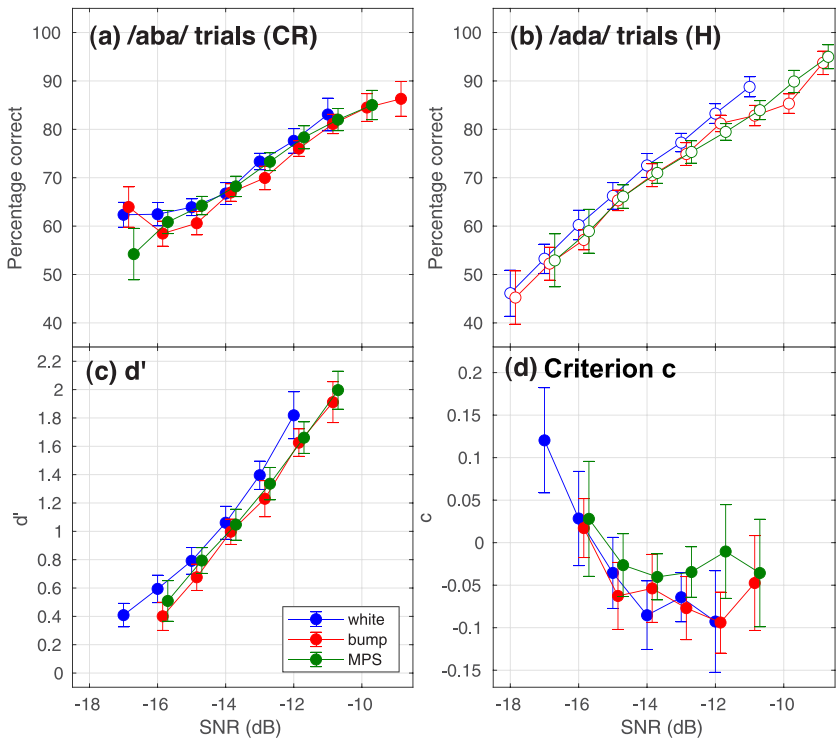

FIG. 5. (Color online) Percentage of correct responses for (a) /aba/ and (b) /ada/ trials, (c) discriminability index $d'$, and (d) criterion, as a function of 1-dB wide SNR bins. These metrics were obtained using the group data. Error bars indicate 1 SEM.

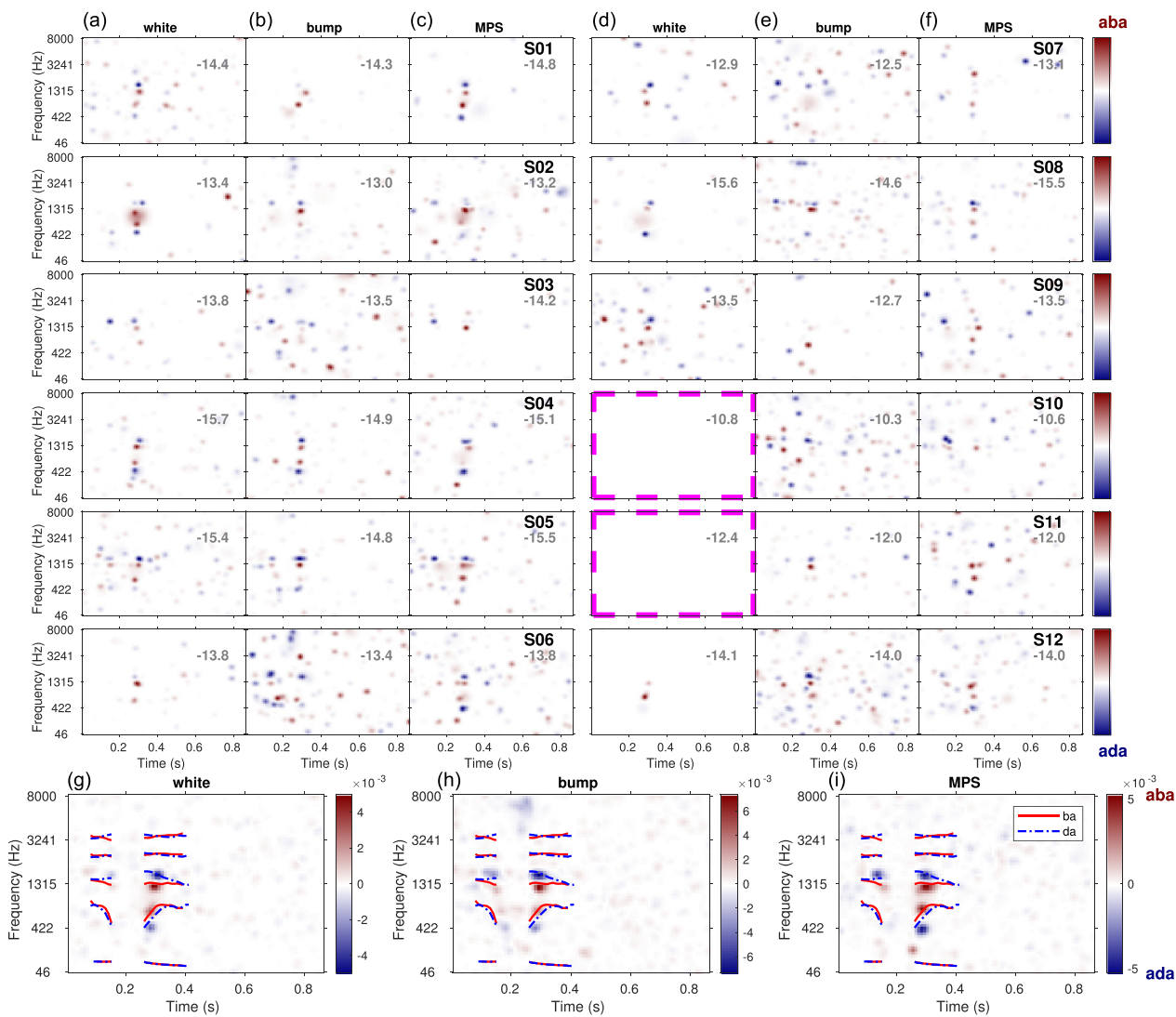

FIG. 6. (Color online) Top panels: ACIs for the 12 participants using (a), (d) white; (b), (e) bump; and (c), (f) MPS noises. For comparison purposes, the weights in each ACI are normalized to their maximum absolute value. The values in gray in the top right corner of each ACI indicate the corresponding mean SNR threshold expressed in dB. The dashed pink boxes indicate the two ACIs that only contain zero weights. Bottom-most panels (g)–(i): mean ACI across all participants, in each condition. The formant trajectories for /aba/ (red solid lines) and /ada/ (blue dotted lines) are superimposed.

When restricting the test set to incorrect trials only [Figs. 7(b), and 7(d)], the prediction accuracy was found to be systematically higher: From the participants' incorrect answers a larger proportion of these errors is explained using the ACIs in bump- ($\Delta CVD_{t,inc} = -3.390 \times 10^{-2}$; $\Delta PA_{inc} = 18.5\%$) and MPS-noise conditions ($\Delta CVD_{t,inc} = -2.920 \times 10^{-2}$; $\Delta PA_{inc} = 17.1\%$) compared to the white-noise condition ($\Delta CVD_{t,inc} = -1.485 \times 10^{-2}$; $\Delta PA_{inc} = 11.3\%$). These group benefits are indicated by filled markers in Fig. 7(d) and are all significantly above chance.

**Cross-predictions between participants**: The procedure to obtain cross-predictions between participants (see Sec. II E 4 b) resulted in three 12-by-12 matrices of cross-prediction values when the data of one participant in one noise condition was predicted using the ACI from another participant in the same condition. The obtained $\Delta PA$ values are shown in Figs. 8(a)–8(c) and ranged between –1.4% and 19.5%. The main diagonal of these matrices correspond to the same auto-prediction values that are shown as open diamond markers in Fig. 7(c). As expected, the $\Delta PA$ values were overall lower (or overall higher using $\Delta CVD_t$) than the auto-prediction values, with on-diagonal averages of 8.1%, 13.3%, and 11.8% for white-, bump-, and MPS-noise conditions, respectively, and corresponding off diagonal averages of 4.7%, 6.0%, and 5.9%. The significance analysis based on $CVD_t$ (only shown in supplementary material Fig. 3, after excluding the predictions marked by red arrows, redrawn in Fig. 8) revealed that 42 (out of 66), 67 (out of 132), and 57 (out of 99) cross-predictions led to a performance that was significantly above chance for white-, bump-, and MPS-noise conditions, respectively. Those cross-predictions are enclosed in pink dashed boxes in supplementary material Fig. 3 and are redrawn in Fig. 8. With this significance analysis, we can identify the ACIs that better predict the data or the data that are better predicted by other ACIs, by looking at the vertical or horizontal direction of the corresponding matrix, respectively. For instance, the ACI from S11 in Fig. 8(c) produced significant predictions using the data of two participants (S01, S05, vertical direction) and the ACI from S05 produced significant predictions using the data of all participants except one (S11). In the cross-prediction of data using other ACIs, the data from S09 for white noise was significantly predicted only using the ACI from S08 [Fig. 8(a), horizontal direction], while the data from S05 in the bump and MPS noise conditions [Figs. 8(b) and 8(c)], were significantly predicted using eight (of 11) other ACIs.

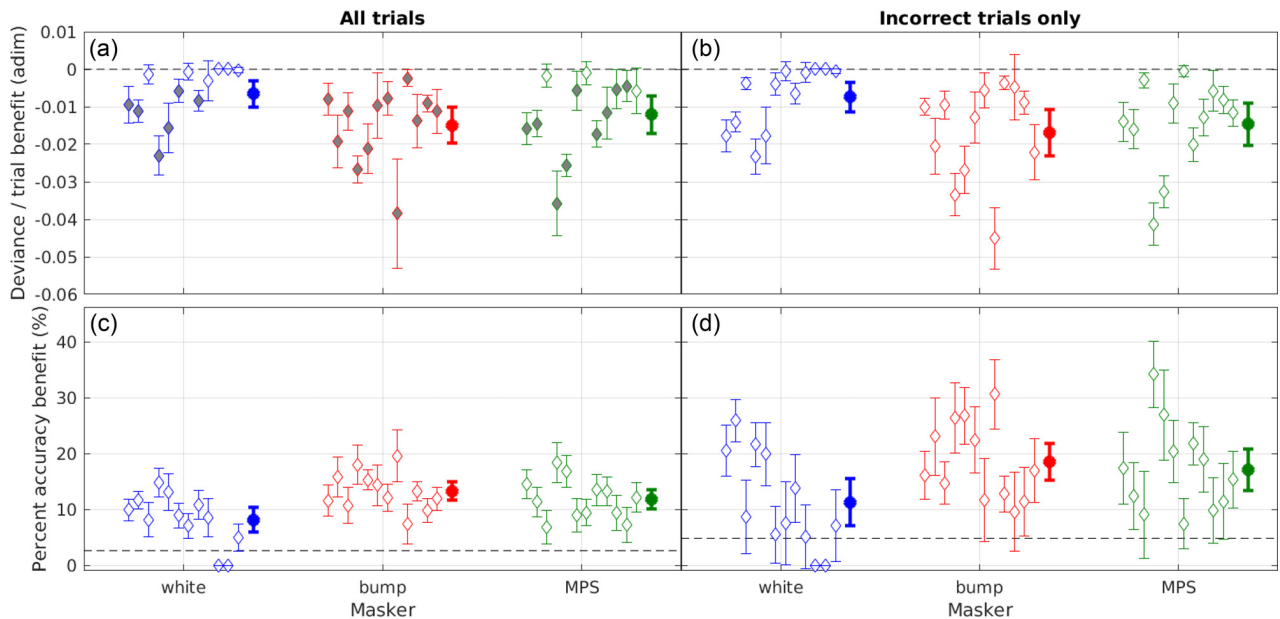

FIG. 7. (Color online) Metrics of performance benefit, $\Delta CVD_t$ (top panels) and $\Delta PA$ (bottom panels), for each of the obtained ACIs, from left to right for participants S01 to S12 (open or gray diamonds). The filled circle markers indicate the group average for the corresponding condition. Left (a), (c) and right (b), (d) panels indicate the analysis using all trials or the incorrect trials only, respectively. In all cases, the error bars indicate ±1.64 SEM and the gray markers in panel (a) indicate those estimates that were found to be significant, based on the boundaries at 0 (dashed line).

**Cross-predictions between noises**: The procedure to obtain cross-predictions between noises but within participant (see Sec. II E 4 b) resulted in twelve individual 3-by-3 matrices, that are shown in supplementary material Fig. 4(A). We focus on the cross-predictions averaged across participants and using $\Delta PA$, which we present in Fig. 9(a). The global results show that the auto (within-noise) predictions of the main diagonals gave overall $\Delta PA$ values of 8.1%, 13.3%, and 11.8% for the white, bump, and MPS noises [the same group values as in Fig. 7(c)] that decreased to (at most) 5.0%, 5.2%, and 5.9%, respectively, when using an ACI to estimate the collected data between noises (compare the elements of Fig. 9 in the vertical direction). Along the horizontal direction, i.e., when exchanged ACIs are used to predict the data within noise condition, the auto-predictions decreased to (at most) 5.2%, 5.0%, and 6.2%, respectively.

### D. Simulations

Simulations were obtained from an artificial listener (Sec. II D and supplementary material Sec. IV), using the same experimental set of noises from participants S01 to S12 and the same methods outlined earlier to derive ACIs. To distinguish the ACIs from the artificial listener from those of the participants, we refer to the first ones as simulated ACIs. The simulated ACIs derived from the waveforms of participants S01–S03 are shown in Fig. 10. The remaining simulated ACIs, are shown in supplementary material Fig. 6.

The simulated ACIs have more clusters of cues, compared to the experimental ACIs (Fig. 6). These clusters seem to be independent of the specific set of noises and are mainly located below 3000 Hz, and between $t = 0.1$ and 0.5 s. All simulated ACIs have large weights in the $F_1$ and $F_2$ regions at $t \approx 0.3$ s.

To quantify the similarity between simulated ACIs we assessed the cross-predictions across "participants," i.e., within noise sets of the same condition. These cross-predictions, using $\Delta PA$, are shown in Figs. 8(d)–8(f). The obtained $\Delta PA$ values were much higher than for the experimental ACIs [panels (a)–(c)], ranging between 33.0% and 43.4% and were always significant. The significant cross-predictions are enclosed by dashed pink boxes and are superimposed to all matrix elements in Figs. 8(d)–8(f)].

The $\Delta PA$ cross-predictions derived from exchanging simulated ACIs across noise conditions for the group are shown in Fig. 9(b). Similar results were found within simulated individuals, as can be found in supplementary material Fig. 4, where all cross-predictions were significantly above chance.

Based on the cross-prediction values averaged across data sets in Fig. 9(b), the auto-predictions were 36.0%, 37.6%, 43.4% in the white-, bump-, and MPS-noise conditions, respectively. The high values of the cross-predictions in the off diagonal, that differ by no more than 4.4% with respect to the on-diagonal values of Fig. 9(b), again, support the similarity among the obtained simulated ACIs.

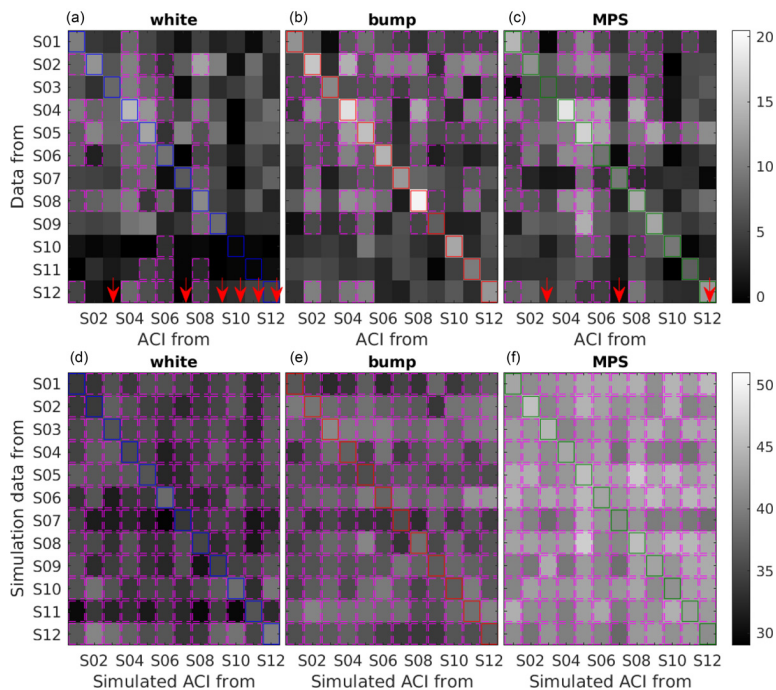

FIG. 8. (Color online) (a)–(c) Between-subject cross-prediction matrices for the three conditions using ΔPA, expressed as corrected-for-chance percentages. The main diagonals are enclosed in colored squares and correspond to the same auto-prediction values as in Fig. 7(c). The pink dashed boxes indicate the ACIs from the abscissa that are able to predict significantly above chance the data of the participant indicated in the ordinate. For this analysis, we only used the ACIs that led to significant auto-predictions [red arrows, related to the gray markers in Fig. 7(a)]. For details, see the text and supplementary material Fig. 3. (d)–(f) Same cross-prediction analysis but using the simulation data. In this case, all cross-predictions led to significant predictions. Note the different (higher) range of ΔPA values with respect to the top panels.

## IV. DISCUSSION

The objective of our study was to measure the token-specific effect of noise on the phonetic discrimination between /aba/ and /ada/. For this purpose, we predicted the listeners' judgements using a microscopic (trial-by-trial) approach, the reverse-correlation method. The results allowed us to derive both macroscopic metrics of speech intelligibility and a microscopic characterization of the participants' listening strategies by means of the time-frequency information in the ACIs.

We start this section by contextualizing the general performance results (Sec. IV A). We then focus on the interpretation of the microscopic ACI analysis to estimate the size of the token-specific effect of noise in our task, going through each of our study hypotheses (Secs. IV B–IV F). We conclude this section by indicating the limitations of the adopted approach (Sec. IV G).

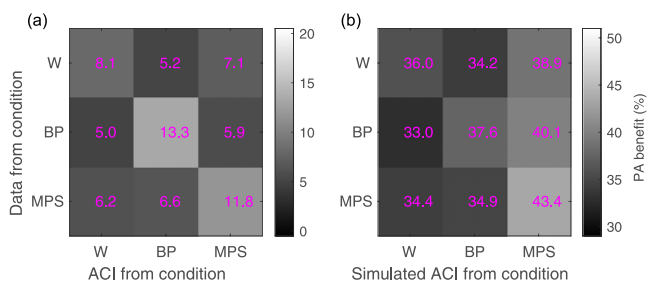

FIG. 9. (Color online) Between-noise cross-predictions using ΔPA values averaged across (a) participants or (b) simulated participants. The off diagonal values were overall lower than the within-noise predictions. See text for further details.

### A. General performance in the task

The behavioral performance for each participant averaged across conditions was very similar, although there were participants with lower or higher overall performance, as indicated by the vertical shift of gray traces in Fig. 4 and as also seen in the SNR thresholds reported in Fig. 6, that ranged between −15.7 and −10.3 dB. When expressing the same data as a function of SNR we observed that, first of all and as expected, the difficulty of the task increased for lower SNRs. More precisely, the percent correct rates for /aba/ [Fig. 5(a)] and /ada/ trials [Fig. 5(b)] decreased from about 90% (for SNRs > − 10 dB) to chance level (for SNRs < − 17 dB). This strong effect of SNR was also visible using the discriminability index $d'$ [Fig. 5(c)] and the criterion $c$ [Fig. 5(d)]. The $d'$ values were lower for bump and MPS noises than for white noises at any SNR. Given that all three noise types have approximately the same long-term spectrum [Fig. 2(b)] and thus should produce a similar energetic masking effect, the lowered discriminability—that leads to an increased number of incorrect answers—can be attributed to the additional random fluctuations between 0 and 30 Hz in the bump and MPS noises [Fig. 2(d), middle and right panels]. Another observation that can be inferred from the lowered discriminability of bump and MPS noises is that there was a marginal, if any, effect of listening in the dips (Cooke, 2006). In fact, a significant effect of this phenomenon should have resulted in better performance for more modulated maskers, an effect that we did not observe. In Fig. 5(d), we observed a bias towards "ada" answers for SNRs below −15 dB, where the criterion values $c$ were higher than 0, in contrast to the nearly constant $c$ values for the SNR bins centered at or above −15 dB.

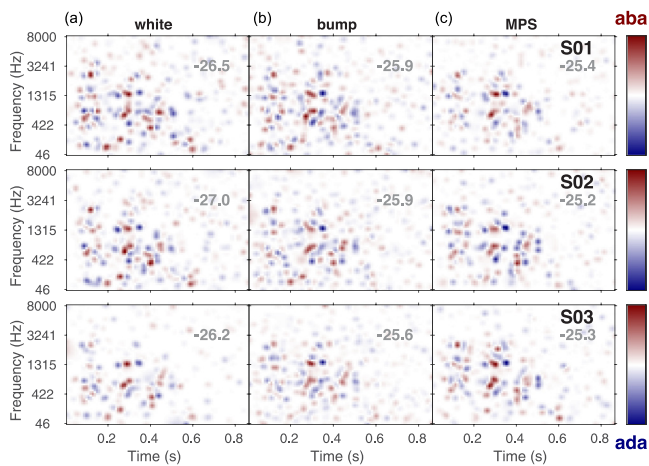

FIG. 10. (Color online) ACIs derived from the simulations using the artificial listener for (a) white, (b) bump, and (c) MPS noises using the set of noises from participants S01–S03 (top to bottom rows). The values in gray indicate the corresponding mean simulated SNR threshold expressed in dB. The ACIs derived from simulations for the remaining set of noises (S04–S12) are shown in supplementary material Fig. 6.

### B. ACIs and token-specific effect

The trial-by-trial (microscopic) analysis based on reverse correlation resulted in a total of 36 ACIs (Fig. 6), that characterized the listening strategy of the twelve participants in each background noise condition. In terms of prediction performance, the obtained ACIs were able to predict the categorical response in each trial ("aba" or "ada") with a better-than-chance accuracy using 27 (out of 36) ACIs [Fig. 7(a), open diamond]. Additionally, the group results were significantly above chance [Fig. 7(a), filled circles], indicating that the exact within-trial noise configuration had a significant influence on the participants' responses or, in other words, that the three types of noises elicited a measurable token-specific effect. This effect was measured to be on average $\Delta PA = 11\%$ across all participants and conditions and ranged between 4.9% and 19.6% for the individual results [open diamonds in Fig. 7, diagonals in Figs. 8(a)–8(c)]. These results agreed with our preregistered expectations about the significance of ACI-based predictions and the size of their effect. Given that the GLM-fitted ACIs only relied on the T-F distribution of random noise envelope fluctuations, without using any explicit information from the targets, these results are supportive of Hypothesis H1 (Sec. I).

The idea that different noise tokens produce a different amount of masking is not new. This has been shown in the context of, e.g., tone-in-noise detection (Ahumada and Lovell, 1971; Pfafflin, 1968), AM-in-noise detection (Varnet and Lorenzi, 2022), but also in psycholinguistic tasks (Varnet et al., 2013; Zaar and Dau, 2015). Using frozen noise, Zaar and Dau (2015) demonstrated that two particular white-noise tokens elicited different confusion patterns. They showed that, when presented with one specific frozen noise token at SNRs below 0 dB, the sound /gi/ was confused with /di/ or /bi/, but when the same sound was presented with a different frozen-noise token, /gi/ was robustly perceived for SNRs down to –15 dB.

A further exploration of the participants' ACIs indicates that the random noise envelope fluctuations that can trigger a token-specific effect are concentrated in small but non-uniformly distributed T-F regions. These regions overlap with the position of acoustic cues in the target sounds, as emphasized in Figs. 6(g)–6(i). For example, the presence of a burst of energy in the random envelope fluctuations in the vicinity of the $F_2$ onset will induce a response bias in favor of "aba" or "ada" depending on whether its spectral position matches the $F_2$ onset frequency of /aba/ (1298 Hz) or /ada/ (1722 Hz). Similarly, subtle differences in the random envelope fluctuations in the region of the $F_1$ onset, the $F_2$ offset in the initial syllable, as well as in the plosive burst, located in the high-frequency region at the consonant onset, affected the listeners' responses in a systematic way. The importance and relative weights of $F_2$-transition cues and burst cues in the perception of voiced plosive consonants has already been discussed in length elsewhere [e.g., Delattre et al. (1955) and Ohde and Stevens (1983)] in particular in the presence of background noise [e.g., Alwan et al. (2011) and Li et al. (2010)]. References to a possible role of $F_1$ transitions are more seldom (Alwan et al., 2011; Delattre et al., 1955) but this cue was already found in our previous ACI studies (Varnet, 2015; Varnet et al., 2015).

As noted above, there seems to be a correspondence between the T-F regions from Fig. 6, where the presence of random envelope fluctuations was particularly detrimental to the listener, and the acoustic cues from the targets. Arguably, this token-specific effect of noise can be seen as the counterpart of the modulation masking effect described in Sec. I, as in both cases random envelope fluctuations induce confusions to the listeners, or a "sorting problem" using Drullman's words. For the modulation masking effect, weak elements of the speech targets are confused with non-relevant noise envelope fluctuations. For the token-specific effect of noise, in contrast, it is the large random noise envelope fluctuations that are confused with relevant elements of the targets if the corresponding T-F locations overlap, affecting the listeners' phonetic decisions. This is reminiscent of the conflicting cues that have been reported from the detailed analyses of phoneme confusions (Li and Allen, 2011; Régnier and Allen, 2008; Singh and Allen, 2012). These studies showed that some speech utterances present incidental acoustic cues that can be confused with characteristic cues of other phonemes, making the targets prone to confusions. In the case of our experiment, the conflicting cues are induced by the background noises.

### C. Token-specific effect in white noise

To estimate the token-specific effect of noise and dissociate it from the overall effect, we used the metrics of out-of-sample prediction accuracy (Sec. II E 4 a) applied to the obtained ACIs. More specifically, we used the prediction accuracy ($\Delta PA$) to measure the size of the effect and the cross-validated deviance per trial ($\Delta CVD_t$) to test its significance. Both metrics were compared to a null ACI, where the weights associated with the T-F noise distributions are set to zero, producing "aba"-"ada" predictions at the participant's chance level.

The contribution of the token-specific effect to the overall effect in white noise was small but significantly above chance for 6 of the 12 participants [Fig. 7(a), blue open markers], with an auto-prediction benefit $\Delta PA$ of 8.1% at the group level [Fig. 7(c), blue filled marker]. This estimate considers trials where conflicting cues misled the participants, as well as trials where the cues reinforced their correct answers. If we restrict this analysis to incorrect trials only, the auto-prediction benefit was significant for 7 participants [Fig. 7(b), blue open markers] with an increased benefit of $\Delta PA_{inc} = 11.3\%$ at the group level [Fig. 7(d), blue filled marker]. This small effect agreed with our preregistered expectations (hypothesis H1) and with the relatively small token-specific effect of noise or "noise-induced effect" reported by Zaar and Dau (2015). Zaar and Dau compared different sources of variability in the recognition of consonant-vowel words presented at different SNRs. In their analysis of source-induced variability they found that the variability in the background noises induced a significant perceptual effect, but that the effect was smaller than the variability in the speech sounds, between- and within-talkers. These observations are in line with our ACI results, although the analysis of Zaar and Dau of within-participant variability was based on confusion matrices that are not directly comparable to our test conditions.

### D. Token-specific effect in more fluctuating noises

The /aba/-/ada/ discrimination using "white-noise-like" bump and MPS noises with emphasized envelope fluctuations, was included to investigate a potential increase in the token-specific noise effect, as they were supposed to mask more efficiently the relevant elements of the speech targets. This is related to our preregistered hypothesis H2 (Sec. I). H2 is supported by the out-of-sample performance results, where $\Delta PA$ increased from 8.1% for white noises to 13.3% and 11.8% in the bump- and MPS-noise conditions, respectively [Fig. 7(c), filled markers] and from $\Delta PA_{inc} = 11.3\%$ to 18.5% and 17.0% for the corresponding analysis with incorrect trials only [Fig. 7(d), filled markers]. At the individual level, all (12 of 12) ACIs produced a significant prediction in the bump-noise condition and 9 of 12 ACIs did so in the MPS-noise condition [Fig. 7(a), red and green open markers]. In other words, with respect to the white noises, bump and MPS noises did not only lead to an increased modulation masking effect as a consequence of their strong envelope fluctuations below 30 Hz [Fig. 2(d)], but they also led to an increased token-specific effect, with respect to the white-noise condition.

### E. Between-subject variability

Due to the nature of our experimental task, that used two vowel-consonant-vowel utterances (/aba/ and /ada/) presented without semantic context, we expected that the ACI method should have resulted in globally similar listening strategies for all our participants. This was related to our preregistered hypothesis H3 (Sec. I). However, in contrast to this expectation, we obtained very heterogeneous ACIs [Figs. 6(a)–6(f)]. We assumed that two ACIs were globally similar when they could predict data significantly above chance when they were exchanged with each other. The results of this analysis were presented in Figs. 8(a)–8(c), where significant cross-predictions using a specific ACI (shown along the abscissa) to predict the participants' data (shown along the ordinate) are enclosed in dashed pink boxes. This analysis revealed that only 42, 67, and 57 (out of 132) cross-predictions led to a performance that was significantly above chance in the white-, bump-, and MPS-noise conditions, respectively. This low number of significant cross-predictions seems to be enough evidence to reject H3.

When analyzing the heterogeneity of the obtained ACIs [Figs. 6(a)–6(f)] we did not find a direct link between the participants' overall performance or any other information about them (e.g., language background, age, or audiometric thresholds) and the exact distribution of obtained T-F cues. For instance, the very good performing participants S04 and S08, who reached SNR thresholds below −15 dB showed significant cross-predictions in all conditions, although the participants differ in their linguistic background. In another example, the bump ACIs from S03 and S10 produced significant cross-predictions despite the difference in overall SNR threshold of 3.2 dB during the experiments, while we did not find a significant similarity in their strategy for the other two noise types.

Differences in listening strategies were also reported by Singh and Allen (2012), who observed a non-negligible between-subject variability in the perception of noisy /b/ and /d/ sounds compared to, e.g., /t/ and /g/. They related this finding to the observation that /b/ and /d/ involve multiple cues (Alwan et al., 2011; Dorman et al., 1977) unlike /t/ and /g/ which have an identifiable single feature that makes them noise robust (Li et al., 2010). Similarly, for our task, the redundancy of available cues noted in Sec. IV B may have enabled our participants to use different listening strategies, as supported by visual inspection of the obtained individual ACIs (Fig. 6). The most logical explanation for such "disparity" in the use of cues may be due to the diversity of listeners' linguistic backgrounds [e.g., Pallier et al. (1997)], however, this contrasts with results of other studies where cue disparities have also been found in participants with a common linguistic background (Clayards, 2018; Singh and Allen, 2012; Zaar and Dau, 2015). Finally, another simple argument could be that the experimental heterogeneity was a side effect of using only two utterances, providing participants with the possibility to focus on more acoustical than phonetic aspects. In any of the cases, further studies are required to investigate the origin of the inter-individual variability observed in the ACIs.

### F. Artificial listener

In this study, the artificial listener was used as a baseline for human performance, under the assumption that measurable changes in auditory-model responses due to changes in the signals—here, using different noise maskers—reflect an effect that might be observable by human listeners (Green and Swets, 1966). More concretely, the artificial listener was used to confirm that (1) a higher out-of-sample prediction is reached for more fluctuating noises (bump and MPS noises) with respect to the steady-state white noises and (2) the specific set of generated noises of the same type does not influence significantly the obtained ACIs. This last point is important because the algorithms for the noise generation were developed and adjusted to particularly influence the modulation frequency content below about 30 Hz and we wanted to confirm that these manipulations do not bias a specific set of responses in the "objective" auditory model decision. At the same time, we expected that the decisions of the artificial listener should elicit a measurable token-specific effect due to the trial-by-trial (microscopic) nature of the template-matching approach.

The results of the simulations were presented in the bottom panels of Figs. 8–9 (out-of-sample metrics), Fig. 10 and supplementary material Fig. 6 (obtained simulated ACIs). All the obtained simulated ACIs produced a measurable token-specific effect of noise with predictions significantly above chance. The token-specific effect using $\Delta$PA, averaged across the 12 data sets, was estimated to be 36.0%, 37.6%, and 43.4% for white, bump, and MPS noises, respectively. These results confirm that a reliable token-specific effect can be measured using the artificial listener and that a mild but systematic increase in $\Delta$PA was observed for the bump and MPS noises with respect to the white noises, in support of hypothesis H4 (Sec. I). Furthermore, the significance analysis showed that all auto-and cross-predictions produced significant out-of-sample metrics [see the pink dashed squares in Fig. 8(d)–8(f)], supporting the statement—also contained in H4—that the specific set of noises did not influence the estimation of simulated ACIs. Despite this support to H4, we observed that the simulated ACIs contain many more T-F cues than the experimental ACIs, suggesting that the underlying strategies between "this" artificial listener and the participants are differently weighted and that, on average, less T-F cues were used by the participants.

### G. Limitations of the approach

The prediction performance of our ACI approach was used to quantify the token-specific effect of noise on an /aba/-/ada/ discrimination, using a single pair of utterances. For this quantification, we assumed that the participants' responses could be predicted purely based on the random envelope fluctuations of the noises that were used to mask the target sounds. Under this strict assumption, we believe that the reported performance metrics represent only a lower bound of the actual token-specific effect because:

(1) We considered the envelope of noise-alone waveforms instead of the envelope of the noisy speech sounds. We assumed this to ensure that the estimated token-specific effect came from the noise maskers and not from the speech targets themselves. Nevertheless, given that envelope extraction is a highly nonlinear process, this assumption implies that the effect of spurious modulations arising from speech-noise interactions (Dubbelboer and Houtgast, 2008; Stone et al., 2011) is negligible.

(2) The transformation of noise waveforms into T-F representations—the inputs to the GLM—is based on a set of linear cochlear filters followed by a simplified envelope extraction (e.g., Osses et al., 2022b), ignoring the potential influence of more central stages of auditory processing on the estimated token-specific effect. In this study, we decided to keep the T-F transformation as simple as possible.

(3) Following a similar principle of simplicity as in the T-F transformation, the GLM approach we used as a statistical model back-end to relate noise (T-F) representations with the participants' responses did not consider interactions between GLM predictors. It is possible, however, that listeners make their decisions based on a non-linear combination of cues.

Another limitation of the present study stems from the use of one single pair of speech utterances. Therefore, one may wonder whether the present results can generalize to other productions of the same logatomes, or to other phonetic contrasts. Some data collected by our group suggest that this is indeed the case. This important aspect will be discussed in a further work.

### V. CONCLUSIONS

In this study, we conducted a microscopic (ACI) analysis of participants' responses in an /aba/-/ada/ discrimination task, using three different white-noise-like and contextless maskers. We demonstrated that:

(1) The detailed noise structure has a measurable effect on a phoneme-in-noise discrimination task. A particular noise token can bias the participants' choice towards one alternative or the other depending on its exact time-frequency (T-F) content. This phenomenon arises from the confusion of random envelope fluctuations in the noise with relevant acoustic cues from the targets.

(2) At low SNRs ($\approx -14$ dB), this effect accounts for at least 8.1% of the participants' responses in white noise (or 11.3% of the errors). When considering other maskers that have larger amounts of random envelope fluctuations, this percentage increased to 13.3% (or 18.5% of errors) and 11.8% (or 17.1% of errors) for the bump and MPS noises, respectively.

(3) Substantially similar results were obtained using an auditory model that is based on a microscopic template-matching approach. The model was used to simulate the same /aba/-/ada/ discrimination task as our study participants. In this case, the token-specific effect of noise was estimated to be between 33.0% and 43.4% of the correct responses. This means that, as expected, the model employed a more efficient and consistent decision strategy, relying on more T-F cues than our participants.

(4) Contrary to hypothesis H3 (Sec. I), we observed a large variability in listening strategies, both between participants and between masker types. A close investigation of the results revealed that, although the primary $F_2$ cue is seen in almost every individual ACI, the weights attributed to secondary acoustic cues appear to differ between participants.

## SUPPLEMENTARY MATERIAL

See the supplementary material for detailed information about (1) the study participants, (2) how to reproduce Figs. 1–10, and (3) replicate our experimental paradigm with human or artificial listeners.

## ACKNOWLEDGMENTS

We would like to thank Richard McWalter for his critical reading of earlier versions of this study and Christian Lorenzi and the rest of the "Modulation Group" for the many discussions during the course of this project. This study was supported by the French National Research agency through the ANR grants "fastACI" (Grant No. ANR-20-CE28-0004) and "FrontCog" (Grant No. ANR-17-EURE-0017).

## AUTHOR DECLARATIONS
### Conflict of Interest

The authors declare they have no conflicts of interest.

## DATA AVAILABILITY

The raw data of this study are available on Zenodo (Osses and Varnet, 2022b). All figures and analyses can be replicated using the fastACI toolbox as of version 1.3 (Osses and Varnet, 2023).

---

[1] Noise maskers are known to degrade target sounds in terms of temporal fine structure [e.g., Drullman (1995)] and temporal envelope. In the present paper we focus on degradations due to temporal envelope information, which are typically assumed to have a larger impact on speech perception [e.g., Dubbelboer and Houtgast (2007) and Shannon et al. (1995)].

[2] Although the choices of the exact supra-threshold SNR and the number of averaged realizations used during the template derivation were arbitrary choices, we checked that they did not significantly affect the simulation results (shown later in Fig. 10). We include a short discussion about these simulation choices in the supplementary material.

[3] We tested different sets of Gaussian-pyramid parameters, exploring the required number of levels and the inclusion of level 0 (i.e., the inclusion of the original $\underline{\underline{N}}_{k,i}$ matrix). The specific configuration of the pyramid did not seem to affect critically the overall shape of the resulting ACIs.

[4] To get an indication of the level of $\Delta$PA that can be attributed to chance only, we assumed that the predictions in each set of 4000 noises follow a binomial distribution $\sim B(4000, 0.5)$, i.e., we assumed that the success of the prediction is determined by chance with $P(r_i = \text{"aba"}) = 0.5$. Considering the one-sided 95% confidence interval, PA needs to be equal to or greater than 51.3% (or $\Delta$PA $\geq$ 2.6%, after correcting for guessing). This boundary is increased to 52.39% ($\Delta$PA $\geq$ 4.78%) for the analysis of incorrect trials, where only 29.3% of the trials are used [$B(1172, 0.5)$].

This "significance test" should only be considered as referential because (1) due to data exclusion, the number of trials is reduced by ≈10%, so 4000 and 1172 are not the exact numbers that should be used in the binomial approximation and (2) the probability of successful prediction by chance deviates slightly from 0.5 depending on the exact ratio of "aba" and "ada" responses in the participant data. To avoid unnecessary confounds, we refrained from including the exact number of trials and chance levels, and we just presented the estimated chance boundary as a visual aid in Fig.7.

Ahumada, A., and Lovell, J. (**1971**). "Stimulus features in signal detection," J. Acoust. Soc. Am. **49**, 1751–1756.

Ahumada, A., Marken, R., and Sandusky, A. (**1975**). "Time and frequency analyses of auditory signal detection," J. Acoust. Soc. Am. **57**, 385–390.

Alwan, A., Jiang, J., and Chen, W. (**2011**). "Perception of place of articulation for plosives and fricatives in noise," Speech Commun. **53**, 195–209.

Clayards, M. (**2018**). "Differences in cue weights for speech perception are correlated for individuals within and across contrasts," J. Acoust Soc. Am. **144**, EL172–EL177.

Cooke, M. (**2006**). "A glimpsing model of speech perception in noise," J. Acoust. Soc. Am. **119**, 1562–1573.

Cooke, M. (**2009**). "Discovering consistent word confusions in noise," in *Proceedings of Interspeech 2009*, pp. 1887–1890.

Dau, T., Kollmeier, B., and Kohlrausch, A. (**1997**). "Modeling auditory processing of amplitude modulation. I. Detection and masking with narrow-band carriers," J. Acoust. Soc. Am. **102**, 2892–2905.

Dau, T., Verhey, J., and Kohlrausch, A. (**1999**). "Intrinsic envelope fluctuations and modulation-detection thresholds for narrow-band noise carriers," J. Acoust. Soc. Am. **106**, 2752–2760.

Delattre, P., Liberman, A., and Cooper, F. (**1955**). "Acoustic loci and transitional cues for consonants," J. Acoust. Soc. Am. **27**, 769–773.

Dorman, M., Studdert-Kennedy, M., and Raphael, L. (**1977**). "Stop-consonant recognition: Release bursts and formant transitions as functionally equivalent, context-dependent cues," Percept. Psychophys. **22**, 109–122.

Drullman, R. (**1995**). "Temporal envelope and fine structure cues for speech intelligibility," J. Acoust. Soc. Am. **97**, 585–592.

Dubbelboer, F., and Houtgast, T. (**2007**). "A detailed study on the effects of noise on speech intelligibility," J. Acoust. Soc. Am. **122**, 2865–2871.

Dubbelboer, F., and Houtgast, T. (**2008**). "The concept of signal-to-noise ratio in the modulation domain and speech intelligibility," J. Acoust. Soc. Am. **124**, 3937–3946.

Elliott, T., and Theunissen, F. (**2009**). "The modulation transfer function for speech intelligibility," PLoS Comput. Biol. **5**, e1000302.

Francart, T., van Wieringen, A., and Wouters, J. (**2011**). "Comparison of fluctuating maskers for speech recognition tests," Int. J. Audiol. **50**, 2–13.

French, N. R., and Steinberg, J. (**1947**). "Factors governing the intelligibility of speech sounds," J. Acoust. Soc. Am. **19**, 90–119.

Glasberg, B., and Moore, B. (**1990**). "Derivation of auditory filter shapes from notched-noise data," Hear. Res. **47**, 103–138.

Green, D. (**1964**). "Consistency of auditory detection judgments," Psychol. Rev. **71**, 392–407.

Green, D., and Swets, J. (**1966**). "Theory of ideal observers," in *Signal Detection Theory and Psychophysics* (Wiley, New York), pp. 151–179.

Harvey, L. (**2004**). *Detection Theory: Sensitivity and Response Bias* (University of Colorado Press, Boulder, CO).

Jørgensen, S., and Dau, T. (**2011**). "Predicting speech intelligibility based on the signal-to-noise envelope power ratio after modulation-frequency selective processing," J. Acoust. Soc. Am. **130**, 1475–1487.

Jürgens, T., and Brand, T. (**2009**). "Microscopic prediction of speech recognition for listeners with normal hearing in noise using an auditory model," J. Acoust. Soc. Am. **126**, 2635–2648.

Kaernbach, C. (**1991**). "Simple adaptive testing with the weighted up-down method," Percept. Psychophys. **49**, 227–229.

Li, F., and Allen, J. (**2011**). "Manipulation of consonants in natural speech," IEEE Trans. Audio. Speech. Lang. Process. **19**, 496–504.

Li, F., Menon, A., and Allen, J. (**2010**). "A psychoacoustic method to find the perceptual cues of stop consonants in natural speech," J. Acoust. Soc. Am. **127**, 2599–2610.

Meyer, B., Jürgens, T., Wesker, T., Brand, T., and Kollmeier, B. (**2010**). "Human phoneme recognition depending on speech-intrinsic variability," J. Acoust. Soc. Am. **128**, 3126–3141.

Mineault, P., Barthelmé, S., and Pack, C. (**2009**). "Improved classification images with sparse priors in a smooth basis," J. Vis. **9**, 17–24.

Noordhoek, I., and Drullman, R. (**1997**). "Effect of reducing temporal intensity modulations on sentence intelligibility," J. Acoust. Soc. Am. **101**, 498–502.

Ohde, R., and Stevens, K. (**1983**). "Effect of burst amplitude on the perception of stop consonant place of articulation," J. Acoust. Soc. Am. **74**, 706–714.

Osses, A., and Kohlrausch, A. (**2021**). "Perceptual similarity between piano notes: Simulations with a template-based perception model," J. Acoust. Soc. Am. **149**, 3534–3552.

Osses, A., Lorenzi, C., and Varnet, L. (**2022a**). "Assessment of individual listening strategies in amplitude-modulation detection and phoneme categorisation tasks," in *International Congress on Acoustics*, hal-03788655, Gyeongju, South Korea, pp. 1–12.

Osses, A., and Varnet, L. (**2021**). "Consonant-in-noise discrimination using an auditory model with different speech-based decision devices," in *Proceedings of DAGA*, hal-03345050, pp. 298–301.

Osses, A., and Varnet, L. (**2022a**). "Auditory reverse correlation on a phoneme-discrimination task: Assessing the effect of different types of background noise," in *ARO Mid-Winter Meeting*, hal-03553443v1.

Osses, A., and Varnet, L. (**2022b**). "Raw and post-processed data for the microscopic investigation of the effect of random envelope fluctuations on phoneme-in-noise perception," doi:10.5281/zenodo.7476407.

Osses, A., and Varnet, L. (**2022c**). "Sound perception using auditory classification images," https://osf.io/4ju3f/ (Last viewed February 12, 2024).

Osses, A., and Varnet, L. (**2023**). "fastACI toolbox: The MATLAB toolbox for investigating auditory perception using reverse correlation (v1.3)," https://github.com/aosses-tue/fastACI (Last viewed February 12, 2024).

Osses, A., Varnet, L., Carney, L., Dau, T., Bruce, I., Verhulst, S., and Majdak, P. (**2022b**). "A comparative study of eight human auditory models of monaural processing," Acta Acust. **6**, 17.

Pallier, C., Christophe, A., and Mehler, J. (**1997**). "Language-specific listening," Trends Cogn. Sci. **1**, 129–132.

Pfafflin, S. (**1968**). "Detection of auditory signal in restricted sets of reproducible noise," J. Acoust. Soc. Am. **43**, 487–490.

Pfafflin, S., and Mathews, M. (**1966**). "Detection of auditory signals in reproducible noise," J. Acoust. Soc. Am. **39**, 340–345.

Plomp, R., and Mimpen, M. (**1979**). "Improving the reliability of testing the speech reception threshold for sentences," Int. J. Audiol. **18**, 43–52.

Průša, Z. (**2017**). "The phase retrieval toolbox," in *AES International Conference on Semantic Audio*, Erlangen, Germany.

Régnier, M., and Allen, J. (**2008**). "A method to identify noise-robust perceptual features: Application for consonant /t/," J. Acoust. Soc. Am. **123**, 2801–2814.

Shannon, R., Zeng, F., Kamath, V., Wygonski, J., and Ekelid, M. (**1995**). "Speech recognition with primarily temporal cues," Science **270**(5234), 303–304.

Singh, R., and Allen, J. (**2012**). "The influence of stop consonants' perceptual features on the articulation index model," J. Acoust. Soc. Am. **131**, 3051–3068.

Stone, M., Füllgrabe, C., Mackinnon, R., and Moore, B. (**2011**). "The importance for speech intelligibility of random fluctuations in steady background noise," J. Acoust Soc. Am. **130**, 2874–2881.

Stone, M., Füllgrabe, C., and Moore, B. (**2012**). "Notionally steady background noise acts primarily as a modulation masker of speech," J. Acoust. Soc. Am. **132**, 317–326.

Varnet, L. (**2015**). "Identification des indices acoustiques utilisés lors de la compréhension de la parole dégradée," Ph.D. thesis, Université Claude Bernard–Lyon I, Lyon, France.

Varnet, L., Knoblauch, K., Meunier, F., and Hoen, M. (**2013**). "Using auditory classification images for the identification of fine acoustic cues used in speech perception," Front. Hum. Neurosci. **7**, 865.

Varnet, L., Knoblauch, K., Serniclaes, W., Meunier, F., and Hoen, M. (**2015**). "A psychophysical imaging method evidencing auditory cue extraction during speech perception: A group analysis of auditory classification images," PLoS One **10**(3), e0118009–23.

Varnet, L., Langlet, C., Lorenzi, C., Lazard, D., and Micheyl, C. (**2019**). "High-frequency sensorineural hearing loss alters cue-weighting strategies for discriminating stop consonants in noise," Trends Hear. **23**, 2331216519886707.

Varnet, L., and Lorenzi, C. (**2022**). "Probing temporal modulation detection in white noise using intrinsic envelope fluctuations: A reverse-correlation study," J. Acoust. Soc. Am. **151**, 1353–1366.

Venezia, J., Hickok, G., and Richards, V. (**2016**). "Auditory bubbles: Efficient classification of the spectrotemporal modulations essential for speech intelligibility," J. Acoust. Soc. Am. **140**, 1072–1088.

Wood, S. (**2017**). "Generalized linear models," in *Generalized Additive Models: An Introduction with R*, 2nd ed. (CRC Press, Boca Raton, FL), Chap. 3, pp. 101–160.

Zaar, J., and Dau, T. (**2015**). "Sources of variability in consonant perception of normal-hearing listeners," J. Acoust. Soc. Am. **138**, 1253–1267.